# Dislocation induced energy dissipation in a tunable trilayer graphene resonator

Lei Yang,[1,2,3] Yifan Huang,[3] Kehai Liu,[3,4] Zhanjun Wu,[1,2,*] Qin Zhou [3,*]

[1] State Key Laboratory of Structural Analysis for Industrial Equipment, School of Aeronautics and Astronautics, Dalian University of Technology, Dalian, China

[2] Key Laboratory of Advanced Technology for Aerospace Vehicles, Liaoning Province, Dalian, China

[3] Department of Mechanical and Materials Engineering, University of Nebraska, Lincoln, Nebraska, USA

[4] Songshan Lake Laboratory for Materials Science, Dongguan, China

*Corresponding authors: wuzhj@dlut.edu.cn (Zhanjun Wu), zhouqin1983@gmail.com (Qin Zhou)

**Abstract**

In crystalline materials, the creation and modulation of dislocations are often associated with plastic deformation and energy dissipation. Here we report a study on the energy dissipation of a trilayer graphene ribbon resonator. The vibration of the ribbon generates cyclic mechanical loading to the graphene ribbon, during which mechanical energy is dissipated as heat. Measuring the quality factor of the graphene resonator provides a way to evaluate the energy dissipation. The graphene ribbon is integrated with silicon micro actuators, allowing its in-plane tension to be finely tuned. As we gradually increased the tension, we observed, in addition to the well-known resonance frequency increase, a large change in the energy dissipation. We propose that the dominating energy dissipation mechanism shifts over three regions. With small applied tension, the graphene is in elastic region, and the major energy dissipation is through graphene edge folding; as the tension increases, dislocations start to develop in the sample to gradually dominate the energy dissipation; finally, at large enough tension, graphene layers become decoupled and start to slide and cause friction, which induces the more severe energy dissipation. The generation and modulation of dislocations are modeled by molecular dynamics calculation and a method to count the energy loss is proposed and compared to the experiment.

During a car crash, the car frame yields and deforms to absorb collision energy and protect passengers. Inside the structural materials, line defects, known as dislocations, emerge and slip to convert part of the mechanical energy into heat [1]. The dislocation dynamics related energy dissipation is important in many plastic deformation processes [2], from stopping a bullet from penetrating a metal plate armor to prevent a bridge from collapsing during an earthquake. The energy dissipation can usually be studied with classic mechanics – for example, by looking at the strain-stress curve during a loading and unloading process of a material slab. If the strain-stress curve during unloading does not retrace the curve during loading, the mechanical energy represented by the area between the loading and unloading curves is lost as heat. However, such study can only analyze the collective effects caused by the dynamics of many dislocations. It is of great interest to directly study the energy dissipation microscopically – for example, how much energy is dissipated during the slipping of a single line of dislocation. The challenge mainly lies in the scale difference: a large number of dislocations could be created, moved, or annihilated when a slab of bulk material undergoes plastic deformation. Isolating just a few is not possible.

In this letter, we report using strain-tunable graphene resonators in the study of dislocation related energy dissipation. Due to graphene's layered structure, dislocations can be created and trapped between as thin as two monolayers [3-5]. This makes few-layer graphene an excellent choice to study the dynamics of dislocations, as well as the related energy dissipation. However, conventional methods are difficult to carry out: the energy dissipation caused by individual dislocations is so small that it is difficult to measure the strain-stress curve accurately in such small and thin samples. Here, we adopt an alternative method to study this problem: by measuring the quality factors (Q-factors) of resonators constructed from few-layer graphene ribbons. On one hand, Q-factors are defined as $Q = 2\pi U/\Delta U$, where $U$ is the vibration energy and $\Delta U$ is the dissipated energy per vibration cycle. On the other hand, Q-factors can also be determined experimentally using the frequency response of resonators (i.e., by measuring the width of the resonance peak). Therefore, measuring Q-factors presents an alternative way to estimate the dissipated energy *ΔU*. Before getting into details, we note that the energy dissipation could include various mechanisms other than dislocation dynamics [6], and needs to be carefully analyzed.

Figure 1 shows the experiment setup (details see supporting information, or SI) that allows us to study the energy dissipation in graphene ribbon in regions where different dissipation mechanism dominates. The different regions are accessed by tuning the strain level in the graphene ribbon using microelectromechanical systems (MEMS). We will show that above a certain strain level, the energy dissipation starts to be dominated by dislocation dynamics. Tuning strain in a graphene resonator has been done in previous works [7-12]. In most works, the tuning is achieved by changing the electrostatic force imposed vertically to the graphene plane [7]. This method is easy to implement, but has limited range of tuning due to the "pull-in" behavior [13]. More importantly, this tensioning method creates strong energy dissipation routes that channels vibration energy (out-of-plane movement, often called flexural modes) to in-plane modes and then to the anchors [14]. This dissipation channel overwhelms others such as the one caused by dislocation dynamics, making the latter difficult to evaluate. A few recent works have demonstrated different strain tuning methods with in-plane tuning force, such as by support shrinkage [8], MEMS actuators [9], and thermally induced tension [10,11]. However, so far, the tunable range of strain is relatively small, and the region where dislocation dynamics dominates energy dissipation has not been achieved. In this work, we constructed a MEMS-graphene hybrid device, where we use silicon based microactuators to apply in-plane strain to a suspended graphene ribbon. The graphene ribbon is driven into resonance, and the relationship among frequency, Q-factor, and strain is carefully investigated. Our fabrication techniques allow us to reliably clamp the graphene ribbons on their ends, therefore a much larger in-plane strain can be achieved, enabling more than 30 folds of resonance frequency tuning before the breakage of graphene ribbons (usually at strain levels of 2~3%).

Fig. 1a is the SEM image of a tunable graphene resonator. Both ends of the suspended graphene ribbon are clamped down by metal patches onto movable silicon structures. The movable structures are connected to microactuators which controls its displacement (Fig. 1b, detailed fabrication process in SI). Electrostatic force is used to drive the vibration of graphene when a DC biased sinusoidal voltage $V_{DC}$ + $V_{AC}\cos(\omega t)$ is applied to the local gate 0.5 µm below graphene (details of experiment setup see SI). We measure the mechanical resonances of graphene by using optical interference microscopy for motion readout at room temperature under vacuum better than $1\times10^{-5}$ Torr. Similar to the ones reported for nanomechanics studies [15-17], our micro actuators are thermally actuated via Joule heating by electrically biasing it. By controlling the actuation voltage applied to the micro actuators, we can control the force applied to the suspended graphene ribbon. To find the strain (in most part of this letter, we are referring to engineering strain, defined as the total sample elongation *ΔL* divided by the sample length *L*) in the graphene sample, an analytical model is developed (See SI for details). We also employ Raman spectroscopy to confirm the actual strain applied (See SI for detailed strain calibration).

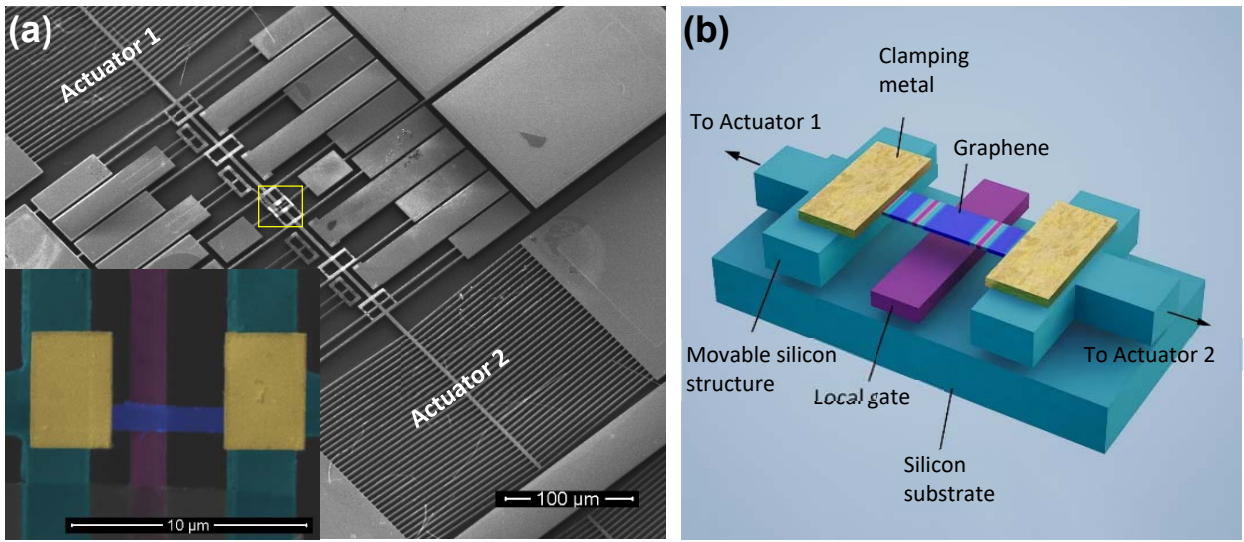

**Figure 1: The tunable graphene ribbon resonator.** (a) SEM images of such a MEMS-graphene hybrid device. The inset is a false-colored magnified view of the center region, with the graphene ribbon in blue color suspended 0.5 µm above the local silicon gate electrode. The schematics is shown in (b) using the same coloring scheme. The in-plane strain in graphene can be tuned by applying tension using the micro actuators, through which dislocations (red lines) can be created.

Fig. 2a presents the frequency tuning of device 1, in which the integrated graphene ribbon is a tri-layer Bernal-stacked sample 0.3-µm wide and 8-µm long. The resonant frequency goes up with increased strain in the graphene ribbon. The tuning behavior can be described by the classical doubly clamped 1-D string model. The resonance frequency of the fundamental mode can be calculated as

$$f_0 = \frac{1}{2L}\sqrt{\frac{T}{\rho}} = \frac{1}{2L}\sqrt{\frac{bE\varepsilon_0}{\rho_0\alpha}} \qquad (1)$$

where, $L$ is the length of the ribbon, $T$ is the tension per unit width, $\rho$ is the area mass density of sheet, $b$ is the thickness (monolayer graphene $b$=0.34 nm), $E$ is the elastic modulus (graphene $E$ = 1 TPa), $\varepsilon_0$ is the in-plane strain, $\rho_0$ is the area mass density of intrinsic graphene sheet (monolayer graphene $\rho_0$ = 7.4 × 10$^{-7}$ kg/m$^2$), and $\alpha$ is a coefficient describing residues adsorbed on the graphene sheet so that $\rho = \alpha\rho_0$ (In SI, we also show that $\alpha$ can be reduced by current annealing the graphene sheet). One may notice that Eq. 1 predicts the resonance frequency approaches zero at very small strain, however, the experiment data begins with a non-zero frequency. The discrepancy comes from the fact that the equation assumes zero bending stiffness of the graphene ribbon, while in reality, as observed in a number of previous works [7,9,18,19], the edges of free-standing graphene sheets are usually folded like a scroll, which significantly increased the bending stiffness. We find the folded edges also significantly contribute to the energy dissipation during vibration, which we will discuss next.

It is previously shown that the Q-factors of similar ribbon-shaped resonators made from silicon nitride can be increased by introducing in-plane tensile strain [20-22]. In these devices, increasing tensile strain can increase the effective spring constant and thus the vibration energy $U$; in the meantime, the dissipated energy $\Delta U$ is not affected. Our initial motivation was to check if this Q-factor enhancement can be repeated in graphene resonators. While the dominant energy dissipation mechanism of the silicon nitride resonators is surface related [21], the energy dissipation mechanism in graphene or other two-dimensional-material-based resonators is still an area of active research. For example, some studies attribute the energy dissipation to microscopic corrugations and wrinkles, which serve as long wavelength elastic scatters carrying away energy from the flexural modes [23]. Applying strain could "iron out" the wrinkles [24] and therefore increase Q-factors. However, only a small amount of strain is required to iron out wrinkles; additional energy dissipation channels need to be found after the wrinkles are ironed out. For example, the Q-factors of a doubly clamped graphene ribbon is usually significantly lower than similar-sized drumhead-like graphene resonators without free edges [18]. Clearly, different from ribbon-shaped silicon nitride resonators, edge-related energy dissipation mechanism can dominate in ribbon-shaped graphene resonators. A few works modeled the effect of free edges [25,26], however, the predicted Q-factors are much higher than experiment measured values.

Fig. 2b shows the experimentally determined Q-factor (calculated as $Q = f_0/\Delta f$, where $f_0$ is the resonance frequency, and $\Delta f$ is the frequency width at half maximum of the vibration amplitude) of the graphene resonator at various strain levels. It can be observed that the Q-factor increases with tension at low strain levels (up to approximately 0.8%), and start to fluctuate as the tension is further increased. The strain dependent behavior inspires us to build a model considering a few possible strain dependent energy dissipation mechanisms. Recall that $Q = 2\pi U/\Delta U$. The vibration energy $U$ in segment $dx$ of the ribbon is:

$$U_{\mathrm{dV}} = dxA \int_0^{\delta\varepsilon} \sigma d\varepsilon = dxA\left(\sigma_0 \delta\varepsilon + \frac{1}{2}E\delta\varepsilon^2\right) \approx dxAE\varepsilon_0 \delta\varepsilon \qquad (2)$$

where $A$ is the cross-sectional area of the graphene ribbon, $\sigma_0$ is the stress level that graphene ribbon is pre-tensioned to, and $\delta\varepsilon$ is the additional strain induced by vibration. $\delta\varepsilon$ can be calculated by noticing the unequal transversal displacement *u(x)* at the left and right end of the element *dx* (small vibration amplitude estimation):

$$\delta\varepsilon(x) \approx \frac{1}{2}\left(\frac{\partial u}{\partial x}\right)^2 \qquad (3)$$

Substitute Eq. 3 into Eq. 2 and integrate over the length of the entire graphene ribbon we obtain

$$U \approx \frac{1}{2}AE\varepsilon_0 \int_0^L dx \left(\frac{\partial u}{\partial x}\right)^2 \qquad (4)$$

which indicates that the vibration energy is proportional to the second power of the vibration amplitude ($U \propto u^2$, through the term $\partial u/\partial x$).

Now consider the dissipated energy *ΔU*. Experimentally, we find that the quality factor has a negligible dependence on the vibration amplitude (unless the resonator is driven to the nonlinear region); therefore the dissipated energy *ΔU* should also be proportional to $u^2$. The dominant energy dissipation mechanism is discussed below. First, we can safely neglect support losses [27] and thermoelastic damping [28] that commonly dominates in silicon-based micro resonators (e.g. the silicon nitride ribbon resonators) since they both predict Q-factors several orders of magnitude larger than the ones observed. In the low strain region, the dominate energy dissipation channel is likely related to the free-edges of the graphene ribbon, since experimentally it is observed that graphene ribbons show much lower Q-factors than drumhead-like graphene membranes without free edges [18]. The exact damping mechanism is still being debated. For example, a few works [25,26] suggest that free edge vibrations will flip between a pair of doubly degenerate warping states, breaking the coherence of the mechanical oscillation of the resonator. However, the Q-factors predicted from this mechanism are much higher than the experimentally measured values [7,18], which suggests this may not be the major damping mechanism. On the other hand, in resonators made of doubly-clamped silicon or silicon nitride ribbons, free edges are not observed to be the major energy dissipation channels [29]. The free-edge induced damping is a distinctive feature of graphene.

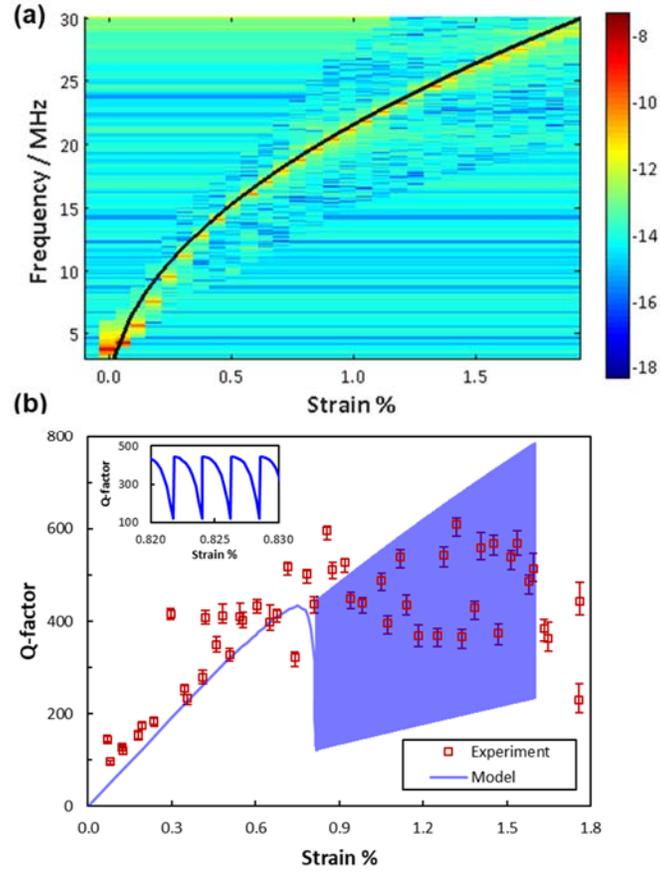

**Figure 2. The tuning behavior of a tri-layer graphene ribbon.** (a) As the in-plane strain in graphene increases from 0 to near 2%, the resonance frequency is tuned from 3.9 MHz to 29 MHz. The black solid line corresponds to the fitted result from Eq. 1. (b) The Q-factor first improves from 100 to 600 at approximately 0.8% strain, but then stops increasing further. The line plots show the results from the energy dissipation models that we proposed. When strain is low the energy loss is mainly from folding edges (Eq. 7); and when the strain is high the energy loss is mainly from dislocations (Eq. 8 with $\eta$ = 0.5 for both developing and developed dislocations).

In Fig. 3, we propose a new mechanism that dominates the damping at low strain levels, based on the alternating folding-and-unfolding of the folded free edges. As mentioned above, the free edges of graphene are often folded into a scroll shape, likely due to surface adsorbent induced stresses (the folding can also be inferred from the frequency tuning data). The folded area *S*, however, changes with the applied tension. This is similar to spinning whirligigs, where the applied tension helps to untwist the twisted ropes. In one vibration cycle of the graphene ribbon, the tension oscillates as the ribbon vibrates; therefore, the folded area *S* is also modulated by the vibration. Since van der Waals contacts are created (destroyed) during the folding (unfolding) of the free edges, energy loss will happen. Next, we try to quantify this energy loss. First, we build a finite element model (details in SI) of the graphene ribbon with a surface absorbent layer on top to provide stresses that fold the free edges. We then apply tension to the ribbon and record the change of the folded area *S* vs. the applied tension (expressed in the uniaxial strain *ε* in the direction of the applied tension). The calculation shows that the rate of area change $\gamma = \partial S / \partial \varepsilon$ is almost a constant in the strain range we run our test. For example, for device 1, we find $\gamma \approx 0.42$ μm². Now to make an estimation on the energy loss during the folding-and-unfolding process, we assume that the van der Waals potential energy is completely lost as heat in the newly folded area in each vibration cycle. This is a reasonable estimation because when making contact, the van der Waals force accelerates the two sides of contact against each other, and the collision between the two sides create lattice vibrations (phonons) that do not contribute to the fundamental vibration mode of the graphene ribbon (i.e. the collision convert van der Waals potential into heat). We can therefore estimate the dissipated energy in segment *dx* of the ribbon per vibration cycle as

$$\Delta U_{edge,dV} = 2\zeta \frac{\gamma}{L} dx \, \delta\varepsilon \tag{5}$$

where $\zeta$ = 0.278 J/m² is the van der Waals binding energy between two graphene layers [30]. The factor of two comes from in one vibration cycle the strain will oscillate twice. Combining Eq. 3 and 5, and integrate over the length of the graphene ribbon we have

$$\Delta U_{edge} \approx \frac{\zeta \gamma}{L} \int_0^L \left(\frac{\partial u}{\partial x}\right)^2 dx \tag{6}$$

Here we can see that the damped energy through this edge mechanism is also proportional to $u^2$. As a result, the quality factor becomes independent of the vibration amplitude and linearly increases with strain:

$$Q_{edge} = \frac{2\pi U}{\Delta U_{edge}} \approx \frac{\pi V E \varepsilon_0}{\gamma \zeta} \tag{7}$$

The calculated Q matches well with the experimental data (Fig. 2a when $\varepsilon < 0.8\%$). Beyond a critical strain of approximately 0.8%, Q stops increasing linearly with strain, indicating a new damping mechanism comes into play.

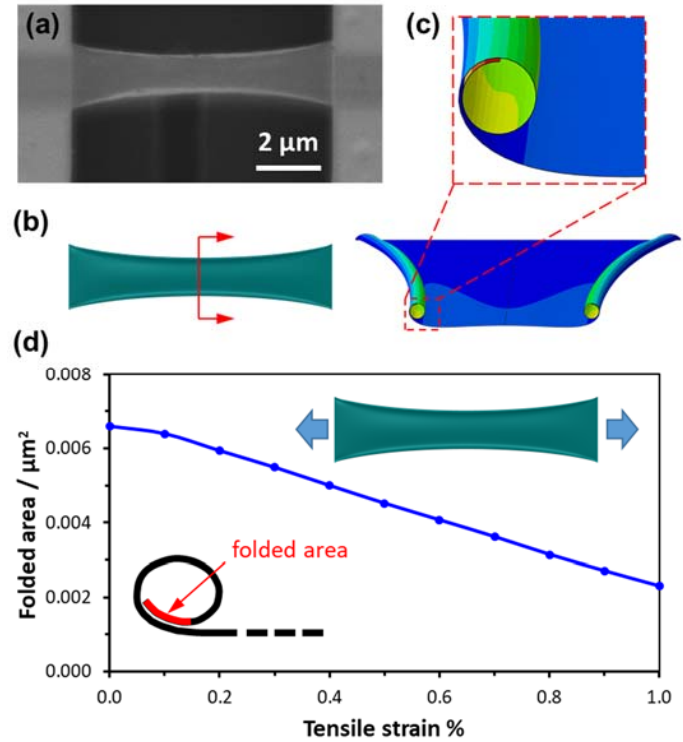

**Figure 3: Energy dissipation through folded edges.** (a) The free edges of a suspended graphene ribbon are often folded as seen in the SEM image. (b)The edge folding, commonly attributed to the stress imbalance from surface adsorbents, are well-captured using finite element analysis. (c) Half-structure and partial enlarged drawing show the detail of the folded edges. The result is color-coded by amount of deformation. With the help of finite element analysis, the relation between the folded area and the externally applied tension on the graphene ribbon can be established. (d) The folded area vs. tensile strain of device 1. As the tension on a graphene ribbon oscillates during vibration, the folded area also varies, resulting in energy dissipation.

Figure 4 illustrates the proposed new damping mechanism that emerges at higher strain levels. Previous works have shown that, when each layer in a multilayer graphene stack is not equally tensioned, dislocations start to appear as strain levels increase beyond some critical value $\varepsilon_c$ [3,31,32]. For example, in bilayer graphene with uniaxial tension only applied to one layer, $\varepsilon_c \approx 0.6\%$ [31]. Similarly, dislocations are likely to develop in the tri-layer graphene in device 1 since layers are not equally tensioned, with the middle layer not subject to external force, while the outer layers do (in contact with the silicon support / clamping metal patches). With dislocations created, the vibration of the resonating graphene ribbon will perturb the strain and consequently modulate these dislocations. Since the stacking energy at dislocations is different from Bernal stacking [5], this modulation can cause energy dissipation if part of the stacking energy change is converted to heat. Here we perform molecular dynamics (MD) simulation (setup details in SI) to estimate the energy dissipation and compare it with our experiments. Due to limitation in computational resources, the simulation is performed on a tri-layer Bernal-stacked ribbon with length of 200 nm. On the right-hand side, the external tension force is applied to the top layer (not to both the top and bottom layers, for reasons explained in SI) from a distance of 50 nm to the right edge, to approximate the clamping metal patches. On the left-hand side, we adopt symmetrical boundary condition (the edges of all layers are fixed in the x-direction), equivalent to stretching a ribbon twice as long from both sides.

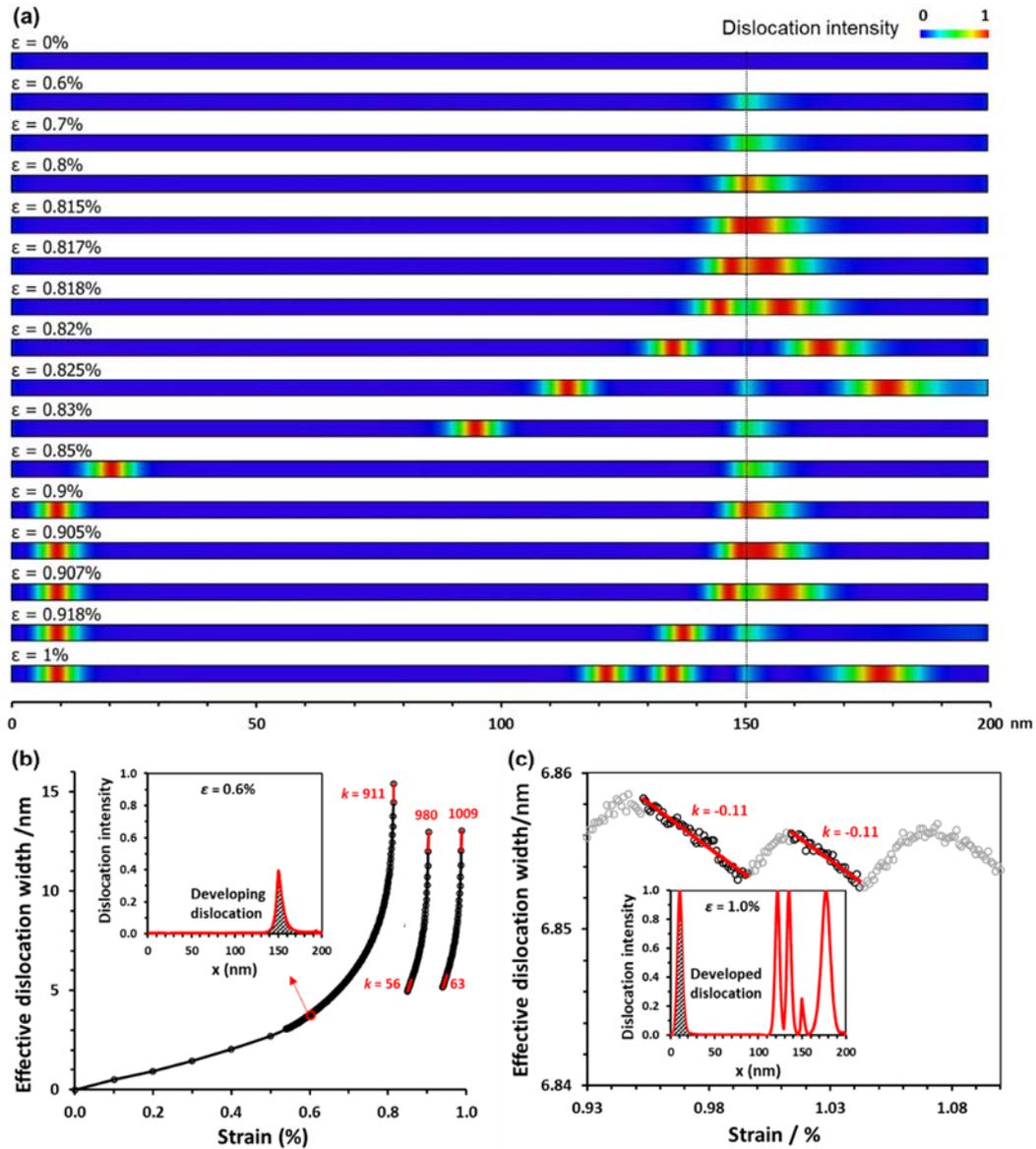

**Figure 4: Energy dissipation through dislocations.** (a) MD simulation results of the formation and development process of dislocations under different applied strains. Here the coloring is the dislocation intensity (DI, definition see SI). (b) Effective width of the developing dislocation as modulated by applied strain. The effective width is calculated using the DI curve (details see SI), proportional to the area of the shadowed region in the inset ($\varepsilon$=0.6%). (c) Effective width of the developed dislocation as modulated by applied strain. Inset: the DI curve of the sample when $\varepsilon$=1.0%.

As shown in Fig. 4a, with the increase of applied strain, a dislocation starts to develop between the clamped (top) and the unclamped (middle and bottom) layers. The width of the developing dislocation grows with applied strain, and it turns into a fully developed dislocation when strain reaches around $\varepsilon_c$ = 0.815%. When this happens, it suddenly splits into two separate dislocations, which move toward left and right respectively. The two fully-developed dislocations continue to move under increased tension, with the left one slows down to a halt (towards the actual sample's center, due to the symmetrical boundary condition) and the right one eventually exits through the right-side edge. Since the right-side developed dislocations does not experience strain modulation during ribbon vibration (top clamped by metal patches), below we focus on the left-side developed dislocations for energy dissipation study. If the strain is further increased, more dislocations will be created in a similar fashion. When a dislocation is created, it pushes all the existing developed dislocations towards left. The number of dislocations can be estimated by noticing that each dislocation creates an area with high local strain and relax the local strain of the rest (Bernal-stacked) part of the sample back to $\varepsilon_c$, effectively offsetting the sample length by $\delta L_d$. Therefore, the number of developed dislocations can be estimated by dividing the additional elongation length ($\Delta L - \Delta L_c$) by $\delta L_d$. Here $\Delta L_c$ is the critical elongation length at which the first dislocation is created. Our MD simulation finds that $\delta L_d$ = 0.18 nm, which means beyond the critical length $\Delta L_c$, with every additional 0.18 nm elongation, a new dislocation will be generated (left side).

When the graphene ribbon vibrates in resonance, its strain oscillates. The strain oscillation modulates both the developing and developed dislocations, mostly on their widths, not positions. To investigate how this small strain oscillation modulates the dislocations, the effective width of the dislocations is calculated as modulated by applied strain. For developing dislocations, the effective width is calculated from 0 to 1% strain (Fig. 4b), from which we can see the continuous increase of dislocation width followed by a sudden drop due to each spilt of the dislocations. The width modulation can be quantified using the rate of effective dislocation width change over strain ($\partial W/\partial \varepsilon$). The range of variation for developing dislocations reflects the fact that the rate is not a constant – it is at minimum (about 60 nm per % strain) at the beginning of developing a new dislocation, and reaches maximum (about 1000 nm per % strain) towards the completion of developing the dislocation. Fig. 4c shows the width modulation for developed dislocations. The part of the curve with downward slope reveals that the dislocation is "squeezed" narrower with increased strain (i.e., $\partial W/\partial \varepsilon < 0$). The part with upward slope is due to the relaxation of strain: when a new dislocation is created, we effectively create an area with high local strain and relax the strain of the rest (Bernal-stacked) part of the sample back to $\varepsilon_c$ (plastic deformation). This relaxation of strain un-squeezes and widens the existing developed dislocations. However, this strain relaxation does not happen during ribbon vibration – this is because no new dislocation is created under cyclic loading (as we have discussed, once a dislocation is created, it is not reversible). Therefore, here we should estimate the part of the curve with negative slopes. We find $\partial W/\partial \varepsilon \approx -0.11$ nm per % strain. We note that the magnitude of the upward slope is determined by the loading rate of MD simulation. Our MD simulation uses a much higher loading rate than the actual experiment due to limited computational resources. In reality, the loading rate is much lower and the creation of a dislocation takes place almost instantaneously. Therefore, the sections of the curve with positive slopes will be short with large slope magnitudes (i.e. the curve shapes like saw tooth).

From the $\partial W/\partial \varepsilon$ values above, the developing dislocations are modulated more heavily by strain than developed dislocations. However, there will only be one developing dislocation near the loading edge, while more and more developed dislocations emerge with increased strain levels. At some point, the energy dissipated through developed dislocation becomes substantial compared to the developing one (see their comparison plot in SI). The increase in the number of dislocations also explains why the increase of Q factors slows down at higher strain levels.

With a clear understanding of the dislocation dynamics in the graphene ribbon, we now try to estimate the energy dissipation during ribbon vibration (details in SI). First we calculate the stacking energy difference $\Delta U_{dis}$ between the higher strain state when the graphene ribbon vibrates to the maximum displacement and the lower strain state when the graphene ribbon is at the balanced position. The Q-factor due to dislocation modulation is:

$$Q_{dis} = \frac{2\pi U}{\eta \Delta U_{dis}} \tag{8}$$

where the coefficient $\eta$ reflect the fact that not all the stacking energy difference $\Delta U_{dis}$ will be dissipated as heat; part of the energy is conservative. This can be understood by noticing that part of the van der Waals force resulted from the stacking potential provides in-plane elastic restoring force for the resonator.

The total energy dissipation of the device is a combination of the edge folding and the dislocation dynamics. The final Q-factor can be calculated as

$$\frac{1}{Q} = \frac{1}{Q_{edge}} + \frac{1}{Q_{dis}} \tag{9}$$

Fig. 2b compares the results from our model to the experiment measured data ($\eta$ is set as 0.5 for both developing and developed dislocations). Note that due to the oscillating behavior of energy dissipation with respect to strain after dislocations are generated, the calculated Q-factor also oscillates quickly (with an approximate period of strain interval of 0.00225%) with respect to strain. The upper and lower limits correspond to the minimal and maximal $\partial W/\partial \varepsilon$ over strain for the developing dislocation, respectively.

The 8 μm long sample can accommodate a maximum number of roughly 400 dislocations at the strain level of approximately 1.5%. Beyond this strain level, the top and bottom layers of graphene are decoupled (our MD result agrees with previous works [31]). We indeed see that our model fails to give good predictions on experimental data beyond ~1.5% strain, where the Q-factor starts to decrease. After decoupling, the sliding between graphene layers may be more appropriately modeled using friction theories [33-35], which is beyond the scope of this work.

In conclusion, we studied the energy dissipation when the tension applied to a tri-layer graphene ribbon is gradually increased. The energy dissipation is too small to be measured by traditional loading-unloading strain-stress curve method, so instead we study the quality factors of graphene ribbons as doubly-clamped 1D resonators. Our graphene-MEMS hybrid devices can apply tunable in-plane tension to the graphene ribbons while driving the ribbons into mechanical resonance. More than 30 folds of resonant frequency tuning can be achieved and from the resonant frequency we can also deduce the strain applied to the graphene ribbon. The quality factors first increase linearly with strain, for which we proposed an energy dissipation mechanism based on the folding-edge. The quality factors start to oscillate and increase sub-linearly at higher strain levels, which we attribute to a new damping mechanism caused by dislocations. The reported graphene-MEMS system proves its capability to serve as a platform for strain engineering and in-situ mechanical studies of suspended graphene and other low-dimensional materials.

**Acknowledgments**

We acknowledge the financial support from the National Key R&D Program of China (2018YFA0702800) and the China Scholarship Council (201706060032).

**Author Contributions**

Q.Z. and Z.W. conceptualized the work. L.Y. and Y.H. made equal contributions in this work. Y.H. and K.L. carried out the experiment. Theoretical modelling and simulations were done by L.Y. All authors have given approval to the final version of the manuscript.

**Competing financial interests**

The authors declare no competing financial interests.